\documentclass[preprintnumbers, prd, showpacs, floatfix,twocolumn,
preprintnumbers, letterpaper, superscriptaddress]{revtex4}
\usepackage{amsfonts}
\usepackage{amsmath}
\usepackage{amssymb,epsf}
\usepackage{latexsym}
\usepackage{graphicx,epsfig}
\usepackage{amssymb}
\usepackage{subfigure}
\usepackage[colorlinks=true,citecolor=blue,linkcolor=blue,urlcolor=black]{hyperref}
\usepackage{epstopdf}

\begin{document}

\title{Critical behaviour and microscopic structure of charged AdS black holes via an alternative
phase space}
\author{Amin Dehyadegari}
\affiliation{Physics Department and Biruni Observatory, College of Sciences, Shiraz
University, Shiraz 71454, Iran}
\author{Ahmad Sheykhi}
\email{ashykhi@shirazu.ac.ir} \affiliation{Physics Department and
Biruni Observatory, College of Sciences, Shiraz University, Shiraz
71454, Iran}
\affiliation{Research Institute for Astronomy and
Astrophysics of Maragha (RIAAM), P.O. Box 55134-441, Maragha,
Iran}
\author{Afshin Montakhab}
\email{montakhab@shirazu.ac.ir} \affiliation{Physics Department
and Biruni Observatory, College of Sciences, Shiraz University,
Shiraz 71454, Iran}

\begin{abstract}
It has been argued that charged Anti-de Sitter (AdS) black holes
have similar thermodynamic behavior as the Van der Waals fluid
system, provided one treats the cosmological constant as a
thermodynamic variable (pressure) in an extended phase space. In
this paper, we disclose the deep connection between charged AdS
black holes and Van der Waals fluid system without extending the
phase space. We keep the cosmological constant as a
fixed parameter and instead, treat the square of the charge of black hole, $%
Q^2$, as a thermodynamic variable. Therefore, we write the
equation of state as $Q^{2}=Q^{2}(T,\Psi)$ where $\Psi$ (conjugate
of $Q^{2} $) is the inverse of the specific volume, $\Psi=1/v$.
This allows us to complete the analogy of charged AdS black holes
with Van der Waals fluid system and derive the phase transition as
well as critical exponents of the system. We identify a
thermodynamic instability in this new picture with real analogy to
Van der Waals fluid with physically relevant Maxwell
construction. We therefore study the critical behavior of isotherms in $%
Q^2-\Psi$ diagram and deduce all the critical exponents of the system and
determine that the system exhibits a small-large black hole phase transition
at the critical point $(T_c,Q^2_c, \Psi_c)$. This alternative view is
important as one can imagine such a change for a given single black hole i.
e. acquiring charge which induces the phase transition. Finally, we disclose
the microscopic properties of charged AdS black holes by using thermodynamic
geometry. Interestingly, we find that scalar curvature has a gap between
small and large black holes, and this gap becomes exceedingly large as one
moves away from the critical point along the transition line. Therefore, we
are able to attribute the sudden enlargement of the black hole to the strong
repulsive nature of the internal constituents at the phase transition.
\end{abstract}

\maketitle

\section{Introduction}

Inspired by the black hole physics a profound connection between the laws of
thermodynamics and the gravitational systems has been argued to exist. A
pioneering work in this respect was done by Bekenstein and Hawking \cite%
{beken,hawk} who disclosed that the entropy ($S$) and temperature ($T$) of a
black hole satisfy the first law of thermodynamics, $dM=TdS$, where $M$ is
the mass of the black hole. Later, the thermodynamic phase space of black
hole was extended by considering the charge $Q$ and the cosmological
constant $\Lambda $ (pressure $P$) \cite{holography1,holography2, mann,
dehghani} as the thermodynamic variables. By consideration the energy
formation of the thermodynamic system, the authors of Ref. \cite{Enthalpy}
showed that the mass of AdS black hole $M$ is indeed the enthalpy $H$.
Therefore, the first law of black hole thermodynamics was written in the
form $dM\equiv dH=TdS+VdP+\Phi dQ$, where $V$ and $\Phi $ are volume and
electrical potential, respectively.

Phase transition have gained attention as a thermodynamic property
of AdS black holes ever since gravity correspondence was
discovered with thermal field theory. The first study on black
hole phase transitions was done by Hawking and Page \cite{HawkPag}
who demonstrated a certain phase transition in Schwarzschild AdS
black hole. This transition can be interpreted a
confinement-deconfinement phase transition in the dual quark gluon
plasma \cite{Witten}. Recently, authors of \cite{sahay} have shown
that Hawking and Page phase transition can be found by Ruppeiner
geometry. One of the important topics in phase transition is the
critical point (continuous phase transition) because thermodynamic
properties of the system exhibit non-analytic behavior. Such
non-analytic behavior is described in terms of power-laws whose
exponents define the universality class of various systems.
Variation of the electric charge affects the thermodynamic
behavior of black hole and consequently it can lead to critical
phenomena. Authors of Refs. \cite{holography1,holography2}
reportedly showed that a
phase transition occurs between large and small black hole in $Q$\textbf{-}$%
\Phi $ plane. They claimed that this behaviour is similar to Van
der Waals phase transition. However, as we will show in this
paper, the phase transition they studied in
\cite{holography1,holography2} when $Q$ is considered as a
thermodynamic variable, is mathematically problematic and
physically unconventional. Similar studies were also carried out
by treating the cosmological constant as the thermodynamic
pressure in an extended phase
space, with its conjugate variable as volume \cite%
{Dolan1,Dolan2,Dolan3,Dolan4,Dolan5,Dolan6}. By exploring the
behavior of
the pressure $P$\ versus specific volume $v$\ (with fixed charge $Q$ and $%
v=2r_{+}$), the authors of Ref.~\cite{mann} showed the existence of a
continuous and discontinuous phase transition between small and large
charged AdS black holes. This transition is analogous to the Van der Waals
liquid-gas phase transition and belongs to the same universality class. In
this view, the cosmological constant is treated as thermodynamically
equivalent to the pressure of the system. However, in general relativity
(GR) the cosmological constant is usually assumed as a constant related to
the background of AdS geometry. Indeed, the cosmological constant, which is
usually assumed as the zero point energy of the field theory, defines the
background of the spacetime. Therefore, from a physical point of view, it is
difficult to consider the cosmological constant as a pressure of a system
which can take on arbitrary values.

It seems natural to think of variation of charge $Q$ of a black
hole and keep the cosmological constant as a fixed parameter,
since the charge of a black hole is a natural external variable
which can vary. However, previous works
\cite{holography1,holography2}  have considered the energy
differential as $\Phi dQ$ with $\Phi =Q/r_{+}$, with $r_{+}$ the
event horizon radius. They have identified a phase transition and
have studied its associated thermodynamic behavior. As we will
show shortly, such a view of thermodynamic conjugate variables
($Q$ and $\Phi =Q/r_{+}$) which are not mathematically independent
can lead to physically irrelevant quantities such as $(\partial
Q/\partial \Phi )_{T}$ which is supposed to be a thermodynamic
response function, but mathematically ill-defined. In the present
work, we offer an alternative view of such an phase space and
definition of new response function which naturally leads to
physically relevant quantity. As we will show, the critical
behavior indeed occurs in $Q^{2}$-$\Psi$ plane, where $\Psi
=1/2r_{+}$. Thus, the first law of black hole thermodynamics as
well as the Smarr relation are subsequently modified. We identify
a small-large black hole phase transition and obtain the critical
point as well as the critical exponents. Perhaps more
interestingly, by calculating the scalar curvature, we are able to
establish a direct link between microscopic interactions and the
resulting macroscopic phase transition.

The outline of our paper is as follows: in the next section, we
study thermodynamics of charged AdS black holes by replacing term
$\Phi dQ$ with $\Psi dQ^{2}$ in the first law of thermodynamics.
In section III, we investigate the critical behaviour and phase
transition of charged AdS black holes by treating the charge as
the thermodynamic variable and keeping the cosmological constant
(pressure) as a fixed parameter. In section IV, we explore
microscopic properties of charged AdS black holes by applying
thermodynamic geometry towards thermodynamics of the system. In
particular, we focus on studying the kind of intermolecular
interaction along the transition curve for small and large black
holes by using the Ruppeiner geometry. We finish with concluding
remarks in the last section.

\section{Thermodynamics of Charged AdS black holes}

Our starting point is the action of Einstein-Maxwell theory in the
background of AdS spacetime \cite{mann}
\begin{equation}
I=\frac{1}{16\pi }\int d^{4}x\sqrt{-g}\left( \mathcal{R}\text{ }-2\Lambda
-F_{\mu \nu }F^{\mu \nu }\right) ,  \label{Act1}
\end{equation}%
where $F_{\mu \nu }=\partial _{\mu }A_{\nu }-\partial _{\nu }A_{\mu }$ is
the electrodynamics field tensor with gauge potential $A_{\mu }$, $\mathcal{R%
}$ is the Ricci\ scalar, $\Lambda =-3/l^{2}$ is the negative cosmological
constant and $l$ is the AdS radius. The line element of Reissner-Nordstr{%
\"{o}}m (RN)-AdS black holes can be written as
\begin{equation}
ds^{2}=-f(r)dt^{2}+\frac{dr^{2}}{f(r)}+r^{2}d\Omega ^{2},  \label{metric}
\end{equation}%
where $d\Omega ^{2}$ is the metric of $2$-sphere and $f(r)$ is given by
\begin{equation}
f(r)=1-\frac{2M}{r}+\frac{Q^{2}}{r^{2}}+\frac{r^{2}}{l^{2}},  \label{fmetric}
\end{equation}%
where $M$ and $Q$ are, respectively, the Arnowitt-Deser-Misner (ADM) mass
and charge of the black hole. The Maxwell equation also yields the electric
field as $F_{tr}={Q}/{r^{2}}$. The event horizon $r_{+}$, is the largest
root of $f(r_{+})=0$, and hence the mass of black hole is written as
\begin{equation}
M(r_{+})=r_{+}+\frac{Q^{2}}{r_{+}}+\frac{r_{+}^{3}}{l^{2}}.  \label{mm}
\end{equation}%
The Hawking temperature of the RN-AdS black hole on event horizon $r_{+}$
can be calculated as \cite{mann}:
\begin{equation}
T=\frac{f^{\prime }(r_{+})}{4\pi }=\frac{1}{4\pi r_{+}}\left( 1+\frac{%
3r_{+}^{2}}{l^{2}}-\frac{Q^{2}}{r_{+}^{2}}\right) .  \label{temp}
\end{equation}%
The entropy of the charged black hole which is a quarter of the event
horizon area is $S=\pi r_{+}^{2}$. Now, we explore thermodynamics of RN-AdS
black hole in a new phase space. We consider the entropy $S$, square of
charge $Q^{2}$ and negative cosmological constant $\Lambda $ ($P=-\Lambda
/(8\pi )$), as independent variables. Thus, the ADM mass of black hole is
enthalpy\cite{Enthalpy}, ($H=M$) and is obtained as
\begin{equation}
M(S,Q^{2},P)=\frac{1}{6\sqrt{\pi S}}\left( 3S+8PS^{2}+3\pi Q^{2}\right) .
\label{M}
\end{equation}%
The intensive parameters conjugate to $S$, $Q^{2}$ and $P$ are defined by
$T\equiv \left( {\partial M}/{\partial S}\right) _{P,\text{ }Q^{2}},\Psi
\equiv \left( {\partial M}/{\partial Q^{2}}\right) _{S,\text{ }P},V\equiv
\left( {\partial M}/{\partial P}\right) _{S,\text{ }Q^{2}}$, \label%
{cparametr} 
where $T$ is the temperature, $\Psi =1/(2r_{+})$, and the thermodynamics
volume $V$ is obtained via $V=\int 4Sdr_{+}=4\pi r_{+}^{3}/3$. The new
variable $\Psi $ is the inverse of the specific volume $v=2r_{+}$ in the
natural unit where $\l _{p}=1$ \cite{mann}. Therefore, the above
thermodynamic relation satisfies the first law as:
\begin{equation}
dM=TdS+\Psi dQ^{2}+VdP.  \label{FL}
\end{equation}%
From scaling argument, we arrive at the Smarr formula as \cite{Enthalpy}:
\begin{equation}
M=2\left( TS+\Psi Q^{2}-VP\right) .  \label{Msm}
\end{equation}%
Let us compare the first law of thermodynamics we have proposed here in Eq.~(%
\ref{FL}) with the usual first law that has been studied in the literature.
For example, the authors of \cite{mann} proposed the first law of
thermodynamics in an extended phase space in the from
\begin{equation}
dM=TdS+\Phi dQ+VdP,  \label{FLs}
\end{equation}%
where $\Phi =Q/r_{+}$ is the electric potential, measured at infinity with
respect to the horizon. The corresponding Smarr formula is \cite{mann}
\begin{equation}
M=2\left( TS-VP\right) +\Phi Q.  \label{Msm2}
\end{equation}%
It is important to note that we have replaced the usual $\Phi dQ$ term in
the first law with $\Psi dQ^{2}$. The extended phase space associated with $%
P=-\Lambda /(8\pi )$ is still the same. Also, the associated Smarr
formula given in Eqs. (\ref{Msm}) and (\ref{Msm2}) are the same, since $%
2\Psi Q^{2}=Q^{2}/r_{+}=\Phi Q$.

Now, the question we ask is which one of the two sets of
equations, i.e. Eqs. (\ref{FL}) and (\ref{Msm}) or Eqs.
(\ref{FLs}) and (\ref{Msm2}) are more appropriate for
investigation of AdS black hole thermodynamics? We have already
argued that the use of extended phase space ($VdP$) is physically
difficult to justify as it implies arbitrary values of
cosmological constant. However, and more important to our propose
here, our question boils down to what is the appropriate
thermodynamic variable
representing the charge of a AdS black hole, $Q^{2}$ or $Q$%
? A look at Eqs. (\ref{fmetric}), (\ref{mm}), (\ref{temp}) and (\ref%
{M}) shows that the charge of AdS black hole is never represented as $Q$%
but always as $Q^{2}$. However, there is far more fundamental
reason for choosing $Q^{2}$ instead of $Q$. The reason is the
corresponding conjugate variable, $\Phi $ vs.$ \Psi $. The
conjugate thermodynamic variable are supposed to be mathematically
independent variables. For example change of volume leads to \ a
change of pressure through some physical process ($VdP$), or
change of temperature leads to change of entropy ($TdS$). It is
through this independent that physically relevant response
functions are defined like compressibility ($-1/V(\partial
V/\partial P)$) and heat capacity ($T(\partial S/\partial T)$),
which are required to be positive for a stable thermodynamic
system. If one looks at the $Q-\Phi $ space, one immediately
realizes such independence is violated as the conjugate variable
$\Phi \equiv Q/r_{+}$ explicitly depends on
$Q$ itself! This makes the definition of a response function $%
(\partial Q/\partial \Phi )$ physically and mathematically
problematic.

Therefore a change in $\Phi $ might automatically bring a change
in $Q$ without any physical response from the system, making
$\partial Q/\partial \Phi $ meaningless as a physical response
function. In order to see how this can lead to problems (see ref.
\cite{holography1} and \cite{holography2} for example), we have
plotted the
Gibbs free energy as a function of $Q$ in the inset of Fig.\ref%
{qphi} and the corresponding \textquotedblleft unstable" isotherm in $%
Q-\Phi $ diagram. The instability associated with multivaluedness
of $G$ is not removed by the standard Maxwell construction as is
clearly seen in the figure. This is so because both regions of
positive and negative slope $(\partial Q/\partial \Phi )_{T}$
remain even after Maxwell construction. Therefore Maxwell
construction does not lead to stable isotherms. Since the entire
thermodynamics of phase transition is based on identification of
stable and unstable regimes, one can conclude that previous
studies which have been based on such a view lead to suspicious
results.

\begin{figure}[h]
$\epsfxsize=7cm\epsffile{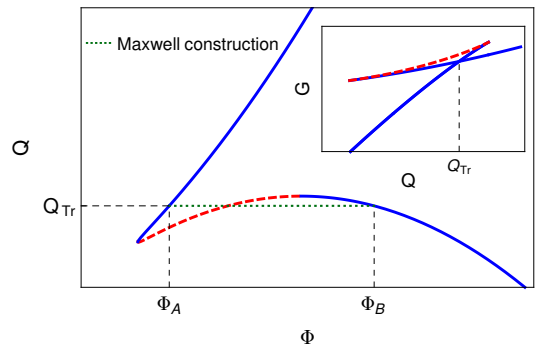}$%
\caption{(Color online) The behaviors of isothermal $Q$-$\Phi $ diagram and
the corresponding $G$-$Q$ diagram (inset) of charged AdS black holes for the
case of $l=1$.}
\label{qphi}
\end{figure}

Here, we propose to consider isotherms in the $Q^{2}-\Psi $
diagrams and use the physically and mathematically relevant
response function $(\partial Q^{2}/\partial \Psi )$ in order to
distinguish regions of stable and unstable system. Note that our
proposed alternative view leads to a natural response function
which measures how the size of a black hole ($\Psi \sim
r_{+}^{-1}$) changes with changes in its charge ($Q^{2}$). We will
see that this alternative view remedies the problems seen
Fig.\ref{qphi} and more importantly leads to a natural
correspondence with the Van der Waals fluid and the associated
small-large black hole phase transition without a need for
extended phase space.

\section{Phase transition and critical exponents}

\begin{figure}[h]
$\epsfxsize=7cm\epsffile{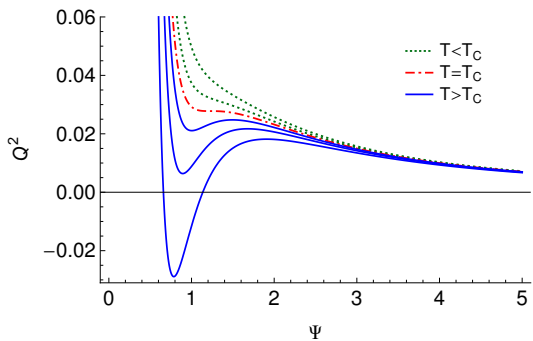}$%
\caption{(Color online) The behavior of isothermal $Q^{2}$-$\Psi $ diagram
of charged AdS black holes for the case of $l=1$.}
\label{qps}
\end{figure}

\begin{figure}[h]
$\epsfxsize=7cm\epsffile{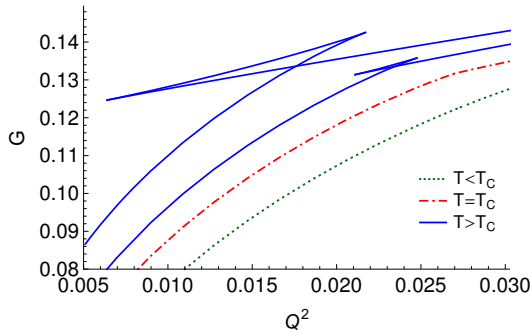}$%
\caption{(Color online) The behavior of isothermal $Q^{2}$-$\Psi $ diagram
of charged AdS black holes constructed by Maxwell equal area law. Here we
have set $l=1$ and rescaled the axes.}
\label{mcon}
\end{figure}

In this section, based on the first law of thermodynamics given in Eq.~(\ref%
{FL}) and the enthalpy (mass) of system given in Eq.~(\ref{Msm}), we propose
an alternative approach towards critical behaviour of black holes by
considering the pressure $P=3/(8\pi \l ^{2})$ as a fixed external parameter
and allow the charge of the black hole to vary. Therefore, by using Eq.~(\ref%
{temp}), one may write the equation of state $Q^{2}(T,\Psi )$ as,
\begin{equation}
Q^{2}=r_{+}^{2}+\frac{3r_{+}^{4}}{l^{2}}-4\pi r_{+}^{3}T.  \label{eqstate}
\end{equation}%
The $Q^{2}-\Psi $ isothermal diagram is shown in Fig. \ref{qps}. In this
figure there are some parts of the isotherms which correspond to a negative $%
Q^{2}$. Clearly these parts of diagram are physically not acceptable. Note
that this also occurs in the usual Van der Waals fluid where the pressure
can become negative for certain values of $T$ \cite{callen}. However, more
importantly, oscillating part of the isotherm indicates instability region ($%
\left( \partial Q^{2}/\partial \Psi \right) _{T}>0$). Both of these
physically unstable features are remedied by the usual Maxwell equal area
construction \cite{callen},
\begin{equation}
\oint \Psi dQ^{2}=0,  \label{ML}
\end{equation}%
as depicted in Fig. \ref{mcon}. Isothermal diagrams show that, for $\l $
constant and $T=T_{c}$, there is an inflection point which is the critical
point where a continuous phase transition occurs. Therefore, the critical
point can be characterized by
\begin{equation}
\frac{\partial Q^{2}}{\partial \Psi }\Big|_{T_{c}}=0,\quad \quad \quad \frac{%
\partial ^{2}Q^{2}}{\partial \Psi ^{2}}\Big|_{T_{c}}=0.  \label{Q1}
\end{equation}%
One obtains:
\begin{equation}
T_{c}=\frac{1}{\pi l}\sqrt{\frac{2}{3}},\quad \quad Q_{c}^{2}=\frac{l^{2}}{36%
},\quad \quad \Psi _{c}=\sqrt{\frac{3}{2l^{2}}},  \label{crivalue}
\end{equation}%
which leads to a universal ($\l $ independent) constant,
\begin{equation}
\rho _{c}\equiv Q_{c}^{2}T_{c}\Psi _{c}=\frac{1}{36\pi }.
\end{equation}%
\bigskip
\begin{figure}[h]
$\epsfxsize=7cm\epsffile{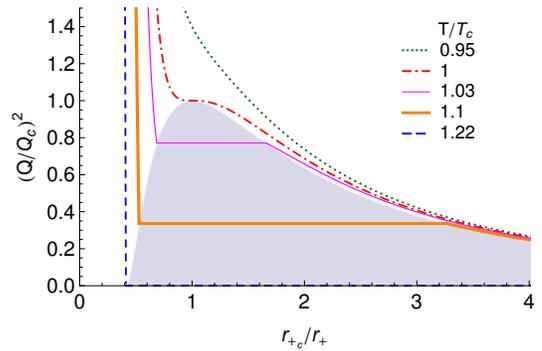}$%
\caption{(Color online) Gibbs free energy of charged AdS black holes with $%
l=1.$ Curves are shifted for clarity.}
\label{gibbs}
\end{figure}
The behavior of a thermodynamic system can also be characterized by the
Gibbs free energy, $G=M-TS.$ In a fixed pressure regime, the Gibbs free
energy reduces to
\begin{equation}
G=G\left( T,Q^{2}\right) =\left( \frac{r_{+}}{4}+\frac{3Q^{2}}{4r_{+}}-\frac{%
r_{+}^{3}}{4l^{2}}\right) \omega ,
\end{equation}%
where $\omega =4\pi $ is the area of unit $2$-sphere, and $r_{+}=r_{+}\left(
T,Q^{2}\right) $, see Eq.~(\ref{eqstate}). The behavior of the Gibbs free
energy in term of $Q^{2}$ is depicted in Fig. \ref{gibbs}. Here, the
multi-valued behavior of Gibbs free energy indicates that the system has a
discontinuous (first order) phase transition from small black hole to large
black hole. Evidently, for $T>T_{c}$ the square of charge and the
temperature of charged AdS black hole are constant during the phase
transition.
\begin{figure}[h]
$\epsfxsize=7cm\epsffile{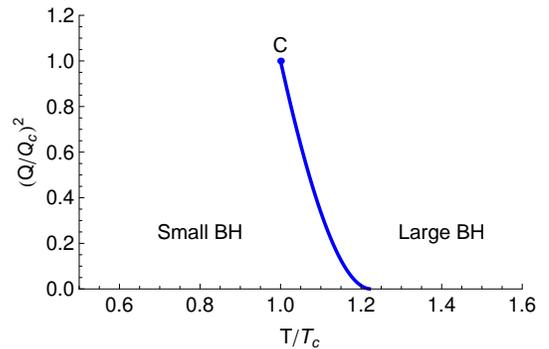}$%
\caption{(Color online) Transition line of small-large BH phase transition
of charged AdS black hole in the $Q^{2}$-$T$ plane. The critical point marks
the end of the transition line.}
\label{qt}
\end{figure}
The transition line can be obtained from Maxwell's equal area law and Gibbs
free energy, which shows a small-large black hole phase transition. Such a
phase diagram is shown in Fig. \ref{qt} where one can see that the extremal
large black hole does not exist.

Next, we turn to calculate the critical exponents in this new \textit{phase
space} approach. The behavior of thermodynamic functions in the vicinity of
the critical point are characterized by the critical exponents. Let us
define the reduced thermodynamic variables
\begin{equation}
\Psi _{r}\equiv \frac{\Psi }{\Psi _{c}},\quad \quad Q_{r}^{2}\equiv \frac{%
Q^{2}}{Q_{c}^{2}},\quad \quad T_{r}\equiv \frac{T}{T_{c}}.  \label{repara}
\end{equation}%
To find the critical exponents, we write the reduced variables in the form $%
T_{r}=1+t,\Psi _{r}=1+\psi ,Q_{r}^{2}=1+\varrho $, where, $t$, $\psi $ and $%
\varrho $ show the deviation from the critical point. First, we consider the
entropy as a function of temperature $T$ and $\Psi =1/(2r_{+})$ as
\begin{equation}
S=S\left( T,\Psi \right) =\frac{\pi }{4\Psi ^{2}}\ ,
\end{equation}%
which is independent of temperature $T$. Therefore the specific heat at
fixed $\Psi $ reads,
\begin{equation*}
C_{\Psi }=T\frac{\partial S}{\partial T}\Big|_{\Psi }=0.
\end{equation*}%
Since, the exponent $\alpha $ describes the behavior of $C_{\Psi }$ near the
critical point as $C_{\Psi }\varpropto \left\vert t\right\vert ^{\alpha },$
one finds $\alpha =0$. By using Eq.~(\ref{repara}), equation of state (\ref%
{eqstate}) translates into the law of corresponding states,
\begin{equation}
Q_{r}^{2}=\frac{6}{\Psi _{r}^{2}}+\frac{3}{\Psi _{r}^{4}}-\frac{8T_{r}}{\Psi
_{r}^{3}},  \label{eqr}
\end{equation}%
which is the equation of state in an $l$-independent form. One can expand
Eq.~(\ref{eqr}) near the critical points as
\begin{equation}
\varrho =-8t+24t\psi -4\psi ^{3}+O\left( t\psi ^{2},\psi ^{4}\right) .
\label{eqexpan}
\end{equation}%
Applying the Maxwell's equal area law and differentiating Eq.~(\ref{eqexpan}%
) with respect to $\psi $ at a fixed $t>0$, leads to
\begin{eqnarray}
\varrho &=&-8t+24t\psi _{l}-4\psi _{l}^{3}=-8t+24t\psi _{s}-4\psi _{s}^{3},
\notag \\
0 &=&\Psi _{c}\int_{\psi _{l}}^{\psi _{s}}\psi \left( 24t-12\psi ^{2}\right)
d\psi ,  \label{maxwell}
\end{eqnarray}%
where $\psi _{s}$ and $\psi _{l}$ denote the event horizon of small and
large black holes, respectively. Eq.~(\ref{maxwell}) have the nontrivial
solution
\begin{equation}
\psi _{s}=-\psi _{l}=\sqrt{6t}.
\end{equation}%
So, the behavior of the order parameter near the critical point can be
calculated as
\begin{equation}
\left\vert \psi _{s}-\psi _{l}\right\vert =2\psi _{s}=2\sqrt{6}%
t^{1/2}\Longrightarrow \beta =1/2.
\end{equation}%
To obtain the critical exponent $\gamma $, we may determine the behavior of
the function
\begin{equation*}
\chi _{_{T}}=\frac{\partial \Psi }{\partial Q^{2}}\Big|_{T},
\end{equation*}%
near the critical point as $\chi _{_{T}}\varpropto \left\vert t\right\vert
^{-\gamma }$. Using Eq.~(\ref{eqexpan}), one obtain
\begin{equation}
\chi _{_{T}}\varpropto \frac{\Psi _{c}}{24Q_{c}^{2}t}\Longrightarrow \gamma
=1.
\end{equation}%
The shape of the critical isotherm $t=0$ is calculated by $\varrho =-4\psi
^{3}\Longrightarrow \delta =3.$ In this way, we have obtained the set of
critical exponents for charged AdS black holes by treating the charge as a
thermodynamic variable and keeping fixed the cosmological constant
(pressure). Note that the obtained critical exponents in this section
coincide with those obtained for Van der Waals fluid \cite{mann,callen}.
Also, the authors of Ref. \cite{mann} obtained the same critical exponents
for charged AdS black holes by keeping the charge as a fixed parameter and
treating the cosmological constant as a thermodynamic pressure and its
conjugate quantity as a thermodynamic volume in an extended phase space. Our
thermodynamic approach to phase space is more natural and potentially more
realistic. For example, one can easily imagine increasing $Q$ while keeping $%
T$ constant and observing the resulting continuous increase in black hole
size $(r_{+})$. More importantly, however, our phase transition diagrams
(e.g. Fig. \ref{mcon}) indicate that there is an instability region where
intermediate-size black hole's should never be observed for $%
T_{c}<T<1.22T_{c}$. This is precisely the Maxwell constructed region where
an abrupt change from a small to large black hole occurs. In the extended
phase space approach this transition occurs as a function of cosmological
constant while in the present approach it occurs as a function of $Q^{2}$.


\section{Thermodynamic Geometry and Microscopic Structure}

\begin{figure}[h]
$\epsfxsize=7cm\epsffile{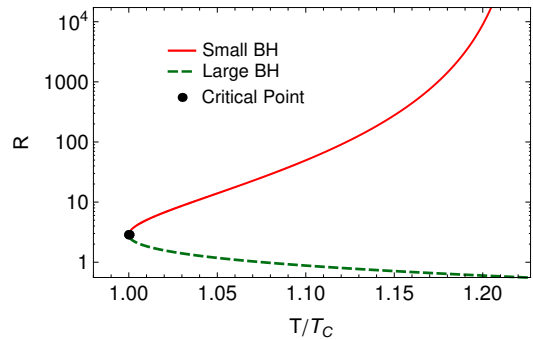}$%
\caption{(Color online) The Ruppeiner scalar curvature $R$ along the
transition curve for small and large black holes. Note the logarithmic scale
on the $R$ axis.}
\label{R}
\end{figure}

In this section we intend to study microscopic properties of charged AdS
black holes by applying thermodynamic geometry towards thermodynamics of the
system. In particular, we focus on studying the kind of intermolecular
interaction along the transition curve for small and large black holes by
using the Ruppeiner geometry obtained from the thermodynamic fluctuation
theory \cite{Rupp1}. Since Ricci scalar is a thermodynamic invariant, the
Ruppeiner geometry defined in ($M$, $Q^{2}$) space can be rewritten in the
Weinhold energy form \cite{Rupp2}:%
\begin{equation}
g_{\mu \nu }=\frac{1}{T}\frac{\partial ^{2}M}{\partial X^{\mu }\partial
X^{\nu }},  \label{GTDmet}
\end{equation}%
where $X^{\mu }=(S,Q^{2})$. We can calculate thermodynamic Ricci scalar
(scalar curvature $R$) that is a thermodynamic invariant analogous to that
of GR. The sign of $R$ gives information about intermolecular interaction in
a thermodynamic system i.e. positivity (negativity) of $R$ refers to
dominance of repulsive (attractive) interaction in thermodynamic system \cite%
{Intraction,comment}. $R=0$ shows there is no interaction in the system \cite%
{idgas}. Using the above, one can calculate the Ruppeiner scalar curvature
with fixed pressure $l=1$. Then%
\begin{equation}
R=\frac{36\left( \Psi /\Psi _{c}\right) ^{2}\left[ \left( \Psi /\Psi
_{c}\right) ^{2}+1\right] }{\pi \left[ 3+6\left( \Psi /\Psi _{c}\right)
^{2}-\left( Q/Q_{c}\right) ^{2}\left( \Psi /\Psi _{c}\right) ^{4}\right] }.
\label{scalar}
\end{equation}%
As shown in Table \ref{tab}, the sign of $R$ depends on the value of $%
Q/Q_{c} $, i.e. the scalar curvature is positive (repulsive intermolecular
interaction) for $\Psi /\Psi _{c}<\sqrt{\Psi _{0}}$ where%
\begin{equation}
\Psi _{0}=\frac{3+\sqrt{9+3\left( Q/Q_{c}\right) ^{2}}}{\left(
Q/Q_{c}\right) ^{2}},
\end{equation}%
and the region of $\Psi /\Psi _{c}>\sqrt{\Psi _{0}}$ is not physically
allowed due to negative absolute temperature.
\begin{table}[h]
\begin{tabular}{c|c|c|}
\cline{2-3}
& $\Psi /\Psi _{c}<\sqrt{\Psi _{0}}$ & $\Psi /\Psi _{c}>\sqrt{\Psi _{0}}$ \\
\hline
\multicolumn{1}{|c|}{$R$} & Positive & Negative \\ \hline
\multicolumn{1}{|c|}{$T$} & Positive & Negative \\ \hline
\multicolumn{1}{|c|}{validity} & allowed & not allowed \\ \hline
\end{tabular}%
\caption{The allowed ranges of $\Psi /\Psi _{c}$.}
\label{tab}
\end{table}
Fig. \ref{R} depicts the behavior of scalar curvature $R$ versus temperature
$T/T_{c}$ along the transition curve (Fig.~4) for small and large black
holes ($T_{c}\leq T<1.22T_{c}$). This figure shows that $R$ is positive
(repulsive intermolecular interaction) and is the same value at critical
point for small and large black hole. Moreover, the Ruppeiner scalar has a
gap between small and large black hole for $T_{c}<T<1.22T_{c}$. Furthermore,
one can note that $R$ becomes increasingly large as $T$ increases for a
small BH (upper branch). This behavior is very interesting as diverging $R$
indicates a very strong repulsive force. For example, our results show that
a small BH at $T\approx 1.22T_{c}$ behave much like a fermi gas at $T\approx
0$\cite{Rupp3}, where fermi exclusion principle dominates the thermodynamic
behavior of the system with strong degenerate pressure.

However, it is also interesting to speculate on the possible microscopic
nature of the phase transition at hand here. As one moves away from the
critical point (continuous transition) along the transition line, one can
see that the corresponding phase transition becomes discontinuous and more
sudden in nature. The resulting large black hole is larger the further away
one moves away from the critical point along the transition line (see
Fig.~2, e.g. $T/T_{c}=1.1$). Now, one can see from Fig.~5 that the
associated $R$ value for such a transition has a larger and larger gap. All
possible stable large back holes have nearly $R$ value of zero, indicating
weakly-interacting constituents. However, the repulsive nature of these
constituents on the small BH side of the transition line can be very large
which is indicative of a very large outward pressure and a tendency to
expand. The picture that emerges here is that phase transition from small to
large black holes is driven by a repulsive nature of the interacting
constituents whose tendencies are to expand the black hole. The larger this
tendency at the transition (i.e. larger $R$) the larger the resulting black
hole (smaller $r_{+}$). We are therefore able to draw certain conclusions
about internal structure and interactions of a black hole simply by looking
at its possible phase transitions.

\section{Concluding Remarks}

In this paper we have studied the small-large black hole phase
transition which have been intensively studied in previous years.
We have not considered the extended phase space view where
cosmological constant is thought to take an arbitrary values. On
the other hand, we have pointed out the problems associated with
the traditional $Q-\Phi $ view in the non-extended phase space.
Subsequently, we have offered an alternative view in the
$Q^{2}-\Psi $ plane where $\Psi =1/2r_{+}$. Unlike previous
studies, this change of view naturally leads to physically and
mathematically meaningful response function $(\partial
Q^{2}/\partial \Psi
)_{T}$ whose sign clearly signifies stable and unstable regimes. %
We have characterized this instability as a small-large black hole phase
transition and have characterized such a transition, including critical
point, critical exponents, and a universal constant. As the name implies,
cosmological constant does not offer a natural variable. However, as we have
shown, for a given cosmological constant, one can imagine black holes taking
on various amount of charge, which can subsequently lead to a phase
transition. We have also studied microscopic properties of charged AdS black
holes by considering thermodynamic geometry. This provides important
insights into the nature of interactions among the black hole's
constituents. Our results indicate that the transition from small black hole
to large black hole is caused by a strongly repulsive interaction amongst
the constituents, which upon expansion and the subsequent phase transition
to a large black hole, lead to a relaxation of such internal forces ($%
R\approx 0$). This is in contrast with the usual liquid-gas transition where
the \emph{attractive} forces among the constituents lead to \emph{%
condensation} and a subsequent large specific volume. We end by observing
that no stable small black hole is possible for $T>1.22T_{c}$, as the
unstable regime extends to arbitrary large $\Psi $, e.g. see Fig.~2. Such
isotherms only exhibit stable large black hole regime with nearly vertical
isotherms. This behavior indicates that, for $T>1.22T_{c}$, the size of the
black hole is essentially a function of temperature only, where addition of
charge does not significantly alter the size of the large black hole as the
constituents have reached a non-interacting regime of $R=0$ and thus no
further expansion is possible. Therefore, in this new alternative view of
the phase space, the attractive force between the constituents is the
deriving force that not only determines the size of the black hole but is
the essential mechanism causing the instability and the subsequent phase
transition.

\acknowledgments {We thank Shiraz University Research Council. The
work of A. Sheykhi has been supported financially by Research
Institute for Astronomy and Astrophysics of Maragha,
Iran.} 

\end{document}